\documentclass[aps,prd,eqsecnum,nofootinbib,floatfix,superscriptaddress,preprint,tightenlines,longbibliography]{revtex4-1}
\usepackage[pdftex]{color}
\usepackage{graphicx}
\usepackage{wrapfig}
\usepackage{bm,amsmath,amssymb}
\usepackage[sort&compress]{natbib}
\usepackage{multirow}
 \usepackage[colorlinks=true,linkcolor=blue,filecolor=blue,urlcolor=blue,citecolor=blue,pdftex,plainpages=false]{hyperref}
\usepackage[left]{lineno}

\begin{document}

\title{Hot-dense Lattice QCD: USQCD whitepaper 2018}

\author{Alexei Bazavov}
\affiliation{Department of Computational Mathematics, Science and Engineering and
Department of Physics and Astronomy, Michigan State University, East Lansing, MI
48824, USA}

\author{Frithjof Karsch}
\affiliation{Fakult\"at f\"ur Physik, Universit\"at Bielefeld, D-33615 Bielefeld,
Germany}
\affiliation{Physics Department, Brookhaven National Laboratory, Upton, NY 11973, USA}

\author{Swagato Mukherjee}
\affiliation{Physics Department, Brookhaven National Laboratory, Upton, NY 11973, USA}

\author{Peter Petreczky}
\affiliation{Physics Department, Brookhaven National Laboratory, Upton, NY 11973, USA}

\collaboration{USQCD Collaboration}
\noaffiliation


\begin{abstract}

  This document is one of a series of whitepapers from the USQCD collaboration. Here,
  we outline the opportunities for, prospects of and challenges to the lattice QCD
  calculations relevant for the understanding of the phases and properties of
  hot-dense QCD matter. This program of lattice QCD calculations is relevant to
  current and upcoming heavy-ion experimental programs at RHIC and LHC.

\end{abstract}

\maketitle

\vspace{3ex}
\centerline{ \textbf{Executive Summary} }
\vspace{3ex}

In 2018, the USQCD collaboration’s Executive Committee organized several
subcommittees to recognize future opportunities and formulate possible goals for
lattice field theory calculations in several physics areas.  The conclusions of these
studies, along with community input, are presented in seven
whitepapers~\cite{Brower:2018qcd,Davoudi:2018qcd,Detmold:2018qcd,Joo:2018qcd,Kronfeld:2018qcd,Lehner:2018qcd}.
This whitepaper provides a roadmap for the current and future lattice QCD
calculations relevant for the understanding of the phases and properties of hot-dense
QCD matter.

The matter that makes up the visible universe is mostly in the form of atomic nuclei.
A nucleus is made up of protons and neutrons, which themselves were shown to be
composed of more basic constituents called quarks, held together by the exchanges of
gluons. The interactions of quarks and gluons are described by the theory of strong
interactions, quantum chromodynamics (QCD). Under extreme conditions of high
temperatures and/or densities hadrons cease to exist; quarks and gluons are liberated
from the hadrons to form a new state of matter, known as the quark-gluon plasma
(QGP).  Understanding the phases of QCD and the properties of QGP  is one of the key
missions of the US nuclear physics program. An entire accelerator-based experimental
facility, the Relativistic Heavy Ion Collider (RHIC) of Brookhaven National
Laboratory, has been devoted to this mission. Many other experimental facilities
across the world, including the Large Hadron Collider (LHC) at CERN, also have joined
this pursuit. The understanding of phases and properties of hot-dense QCD matter from
experimentation, as well as planning of future experiments, need many theoretical
inputs. Lattice-regularized QCD, a technique suited for large-scale numerical
calculations of QCD, is presently the only viable theoretical tool to study QCD in
its full glory, by starting from the fundamental quark-gluon degrees of freedom and
by taking into account the entire complexities of the strong interaction. In light of
the ongoing and future heavy-ion experimental programs at RHIC and LHC, this USQCD
whitepaper outlines the opportunities for, prospects of and challenges to the
hot-dense lattice QCD calculations in addressing the issues: (i) phases and
properties of baryon-rich QCD, (ii) microscopy of QGP using heavy-quark probes, (iii)
nature of QCD phase transitions, (iv) electromagnetic probes of QGP, (v) jet energy
loss in and viscosities of QGP.

\newpage

\section{Introduction}

\begin{wrapfigure}{l}{0.52\textwidth}
  \centering

    \vspace{-4ex}
    \includegraphics[width=0.5\textwidth, height=0.25\textheight]{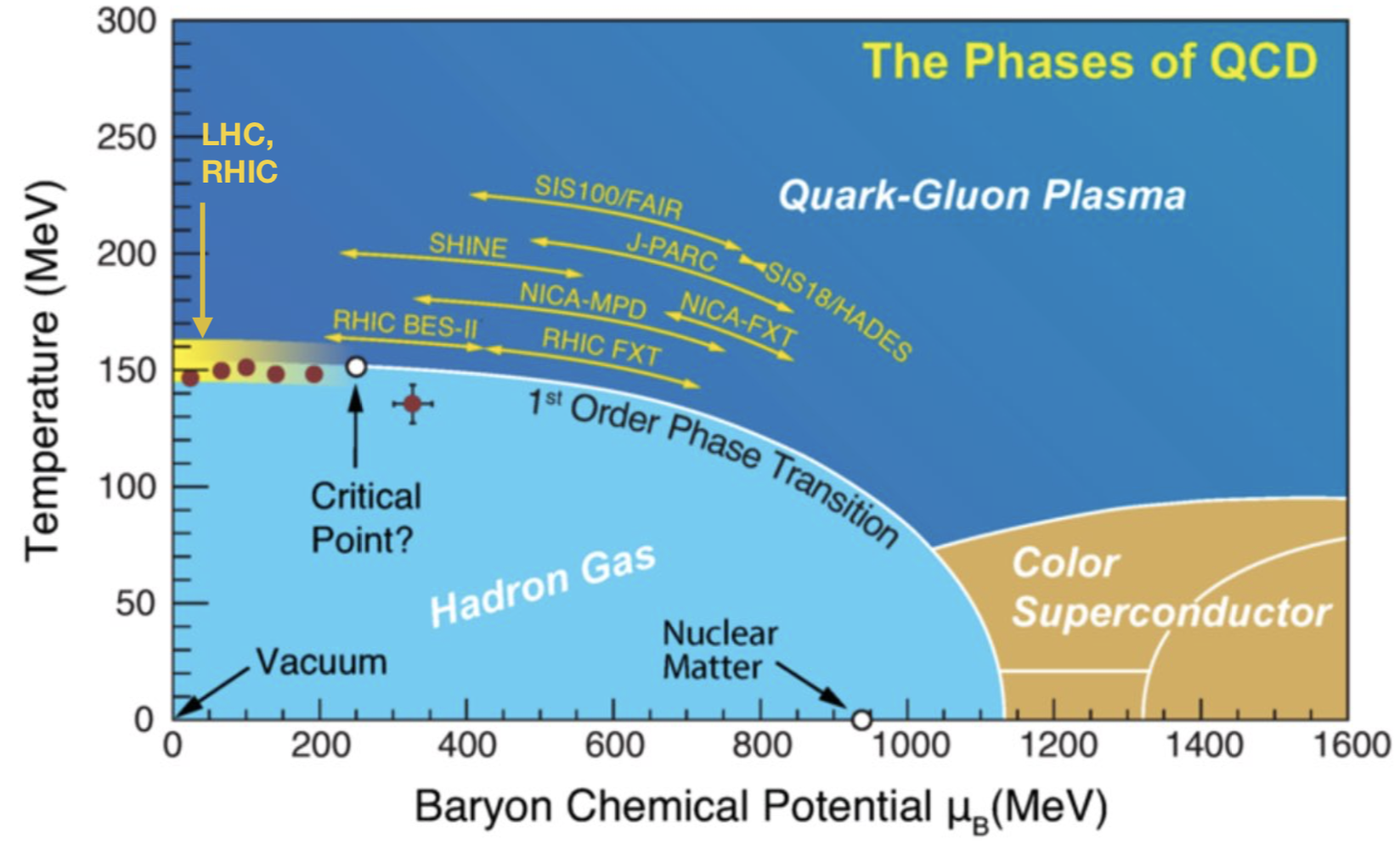}
    \vspace{-1ex}
    \caption{A schematic phase diagram of QCD. Also, indicated are the ranges of
    explorations by various heavy-ion collision experiments.}
    \vspace{-2ex}

\end{wrapfigure}

The mission of the US Department of Energy's (DOE) Nuclear Physics program is to
discover, explore, and understand all forms of nuclear matter. As outlined in the
2015 NSAC Long Range Plan~\cite{Geesaman:2015fha}, a key component of the mission of
this program is mapping the phase structures of quantum chromodynamics (QCD) and
decoding the properties of quark-gluon plasma (QGP). DOE has dedicated an entire
accelerator-based experimental program, the Relativistic Heavy-Ion Collider (RHIC) at
Brookhaven National Laboratory (BNL), in pursuit of these causes. In the Large
Hadron Collider (LHC) at CERN, Switzerland, the entire ALICE detector  is devoted to
exploration of properties and phases of QCD. Lattice-regularized QCD is, presently,
the only viable technique that allows  non-perturbative, parameter-free determinations
of the properties and phase structures of hot-dense strong-interaction
matter from its fundamental theory, QCD. Over the last decade, lattice QCD has proven
to be the most successful technique for model-independent, first-principle
calculations of the phase structures and properties of hot-dense QCD matter; for
recent reviews see Refs~\cite{Ding:2015ona, Soltz:2015ula, Ratti:2018ksb}.

Experimental explorations at RHIC and LHC have revealed the surprising fact that the
long-distance behavior of QGP closely resembles that of an almost inviscid fluid. QGP
created at LHC and top RHIC energies consists of almost as much antimatter as matter,
and is characterized by the nearly vanishing baryon-number chemical potential.  Under
these conditions the transition from the QGP to a hadron gas occurs through a smooth
crossover, with many thermodynamic properties changing dramatically, but
continuously, within a narrow range of temperature~\cite{Bhattacharya:2014ara,
Bazavov:2011nk, Aoki:2006we}. On the other hand, the baryon-rich QGP created at lower
RHIC energies may experience a sharp first-order phase transition as it cools, with
bubbles of QGP and bubbles of hadrons coexisting at a well-defined temperature. This
region of co-existence ends in a critical point, where QGP and ordinary hadron-matter
become indistinguishable.

The experimental explorations of phases of QCD and properties of QGP will continue over,
and beyond, the next decade in many accelerator facilities across the world. For a
recent comprehensive review on the science goals of the future heavy-ion collision
experiments see Ref.~\cite{Busza:2018rrf}. The two central scientific goals
underlying  these experiments are: (i) Explorations of the phases of baryon-rich QCD,
including the search for the QCD critical point; (ii) Understanding the nature of QGP
at shorter and shorter length scales. These themes also are at the heart of the
planned upgrades of the US-based heavy-ion experiments at RHIC. The second phase of
the RHIC Beam Energy Scan (BES-II) program~\cite{Keane:2017kdq}, scheduled for
2019-21, will explore the QCD phase diagram. The sPHENIX
experiment~\cite{Adare:2015kwa} at RHIC, with a planned start in 2023, will probe the
short-distance physics of QGP using bottomonia and jets. Heavy flavor and jet physics
also are key targets of the upgraded ALICE experiments starting 2021, as well as for
the heavy-ion programs of the CMS, ATLAS and LHCb experiments at LHC. Another key
component of the ALICE experiment will be establishing the nature of QCD transition at
vanishing baryon density by looking at the higher moments of conserved charge
fluctuations. Additionally, in future, various electromagnetic probes of QGP will be
studied in detail both at RHIC and LHC. As in the past, success and planing of these
future heavy-ion experimental programs crucially depend on various inputs from
hot-dense lattice QCD calculations. In this USQCD white paper we briefly outline the
hot-dense lattice QCD calculations that will not only enhance our fundamental
understanding of the phases and properties of strong-interaction matter, but also
significantly impact the  heavy-ion collision experiments, particularly the ones at
RHIC.


\section{Phases and properties of baryon-rich QCD}

\begin{wrapfigure}{l}{0.52\textwidth}
    \centering

        \vspace{-6ex}
        \includegraphics[width=0.51\textwidth, height=0.26\textheight]{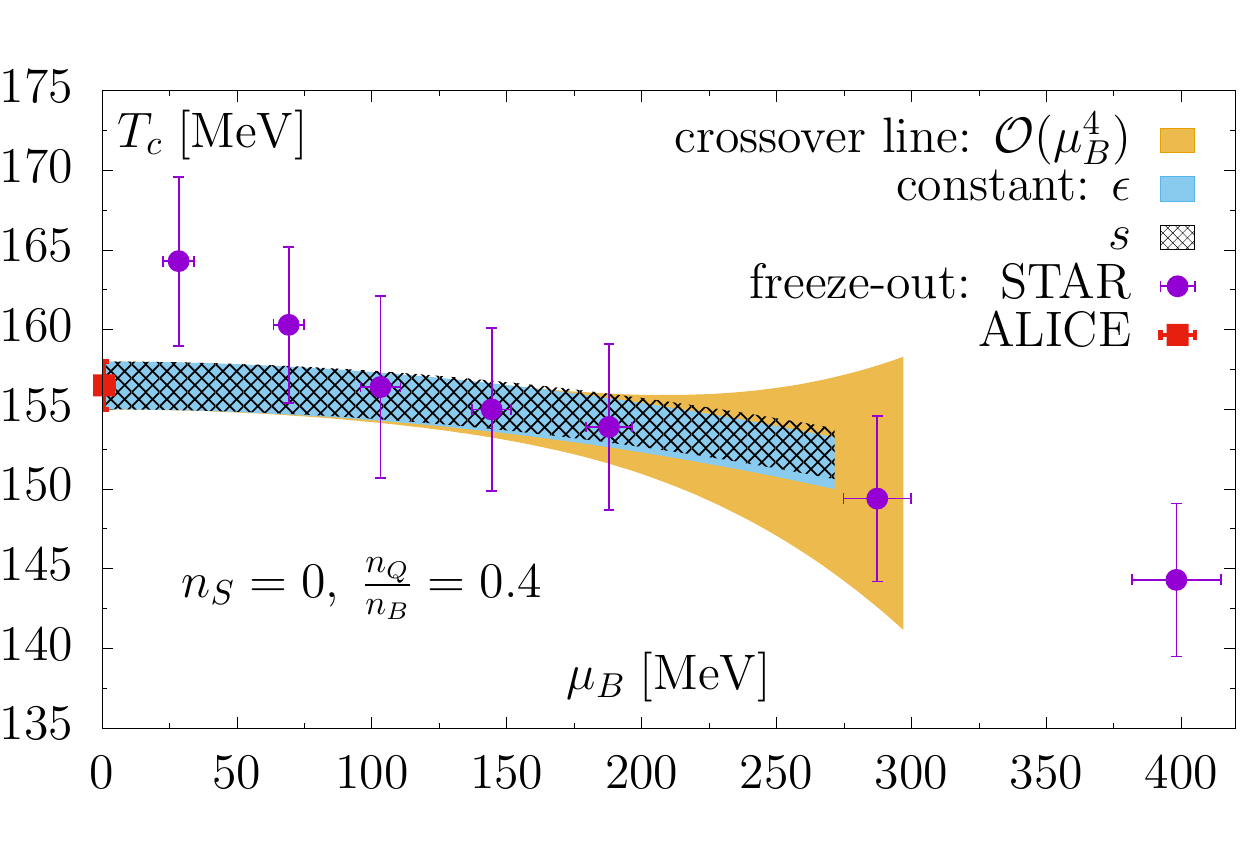}
        \vspace{-3ex}
        \caption{Phase boundary for (2+1)-flavor QCD in the temperature and
        baryon chemical potential plane \cite{Bazavov:2018mes}. Also, shown are lines
        of constant energy and entropy density, as well as the freeze-out
        temperatures determined by STAR at RHIC and ALICE at LHC.}
        \vspace{-2ex}

\end{wrapfigure}

Many properties of strong-interaction matter at non-zero temperature have been
analyzed in hot-dense lattice QCD calculations for vanishing values of chemical
potentials (for recent reviews see, \textit{e.g.}, \cite{Ding:2015ona,
Ratti:2018ksb}). The pseudo-critical (crossover) temperature for the transition from
a low temperature hadronic phase to a high temperature QGP phase has been examined
and extrapolated to the continuum limit for physical values of two degenerate light
(up, down) quark masses and a physical strange quark mass~\cite{Aoki:2009sc,
Bazavov:2011nk}. A recent update of these calculations yields as pseudo-critical
temperature $T_{pc} = (156\pm 1.5)$~MeV~\cite{Bazavov:2018mes,Steinbrecher:2018phh}, which is in
excellent agreement with the freeze-out temperature for hadrons that has been
extracted from particle yields measured by the ALICE collaboration at LHC using a
statistical hadronization model~\cite{Andronic:2017pug}. Also, continuum extrapolated
results for the equation of state at vanishing baryon chemical potential, obtained
with two different staggered fermion discretizations, agree quite
well~\cite{Borsanyi:2013bia,Bazavov:2014pvz,Bazavov:2017dsy}. These results find applications in
hydrodynamic modelings of the expanding matter created in heavy-ion collisions \cite{Auvinen:2018uej} and
statistical analyses of freeze-out conditions.

The current focus of the lattice calculations of bulk properties of
strong-interaction matter concerns the extension of these results to non-zero baryon
chemical potential. As direct lattice QCD simulations are not possible in this case,
because of the notorious sign problem, calculations are done by either using Taylor
expansions~\cite{Gavai:2003mf, Borsanyi:2012cr, Bazavov:2017dus} or by analytical
continuation of results obtained in simulations with imaginary chemical
potentials~\cite{DElia:2002tig, Gunther:2016vcp, Bonati:2018nut}.
Continuum-extrapolated results for the equation of state, now, have been obtained up
to ${\cal O}(\mu_B^6)$ in a Taylor series, as well as through analytic continuations.
This allows one to obtain results for bulk thermodynamic observables, such as the QCD
equation of state~\cite{Borsanyi:2012cr, Bazavov:2017dus}, as well as the curvature
of the pseudo-critical line~\cite{Bonati:2018nut, Bazavov:2018mes}  up to $\mu_B
\simeq (1.5-2)T_{pc}$, which is sufficient as input for the analysis of data from the
RHIC BES down to beam energies $\sqrt{s_{_{NN}}}\simeq 12$~GeV. \emph{In order to
provide input at lower beam energies that will be probed in the upcoming BES II
higher precision for the existing expansion coefficients and results for higher order
terms are needed.}

\begin{wrapfigure}{l}{0.52\textwidth}
    \centering

        \vspace{-3ex}
        \includegraphics[width=0.51\textwidth, height=0.26\textheight]{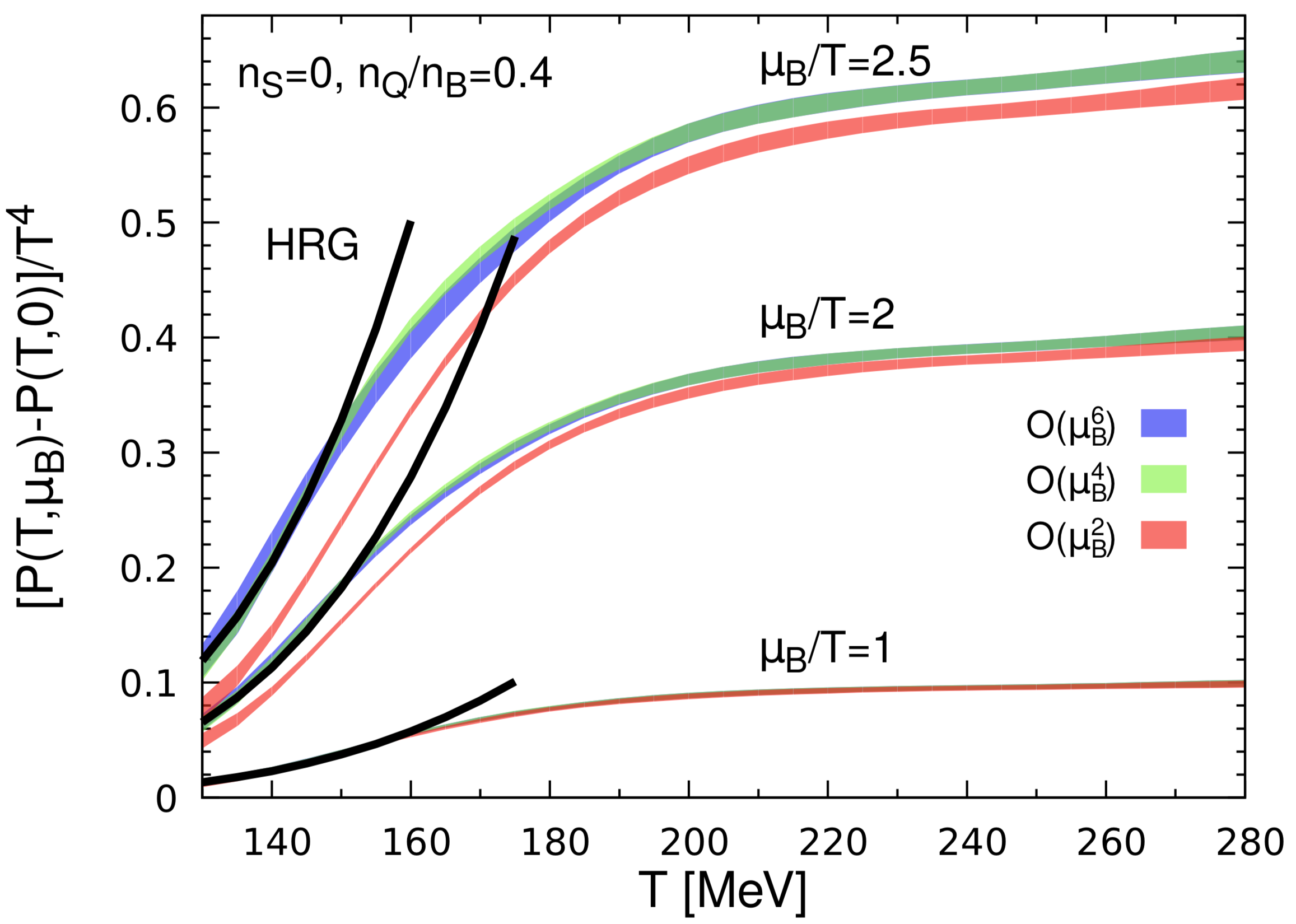}
        \vspace{-1ex}
        \caption{The QCD pressure at non-zero baryon chemical potential for a
        strangeness neutral medium  ($n_S=0$) with net electric charge to baryon
        number density $n_Q/n_B=0.4$~\cite{Bazavov:2017dus}. }
        \vspace{-2ex}

\end{wrapfigure}

Good quantitative control over higher order Taylor expansion coefficients for bulk
thermodynamic observables and fluctuations of conserved charges also is needed to
estimate the convergence of these expansions~\cite{Borsanyi:2018grb}. For
sufficiently high orders in the expansion the systematic of the sign changes in the
expansion coefficients provide estimators for the possible location of a critical
point in the $T-\mu_B$ phase diagram. Here the current estimates are limited by the
statistical accuracy of the higher order expansion coefficients. Current estimates
from up to $8^{th}$ order expansion coefficients suggest that it is unlikely to find
a critical point located at baryon chemical potentials smaller than $\mu_B\sim
2T_{pc}$ and temperatures larger than $T\sim 140$~MeV. This is
consistent with the even smaller value of the chiral phase transition
temperature at vanishing light quark masses and physical
value of the strange quark mass, which for vanishing baryon chemical
potential is found to be
$T_c^0= 132^{+3}_{-6}$ \cite{Ding:2019prx}. \emph{Improving over these
estimates requires calculations at lower temperatures and higher accuracy on at least
10$^{th}$ order Taylor expansion coefficients.}

\begin{wrapfigure}{l}{0.52\textwidth}
  \centering

  \vspace{-3ex}
  \includegraphics[width=0.5\textwidth, height=0.25\textheight]{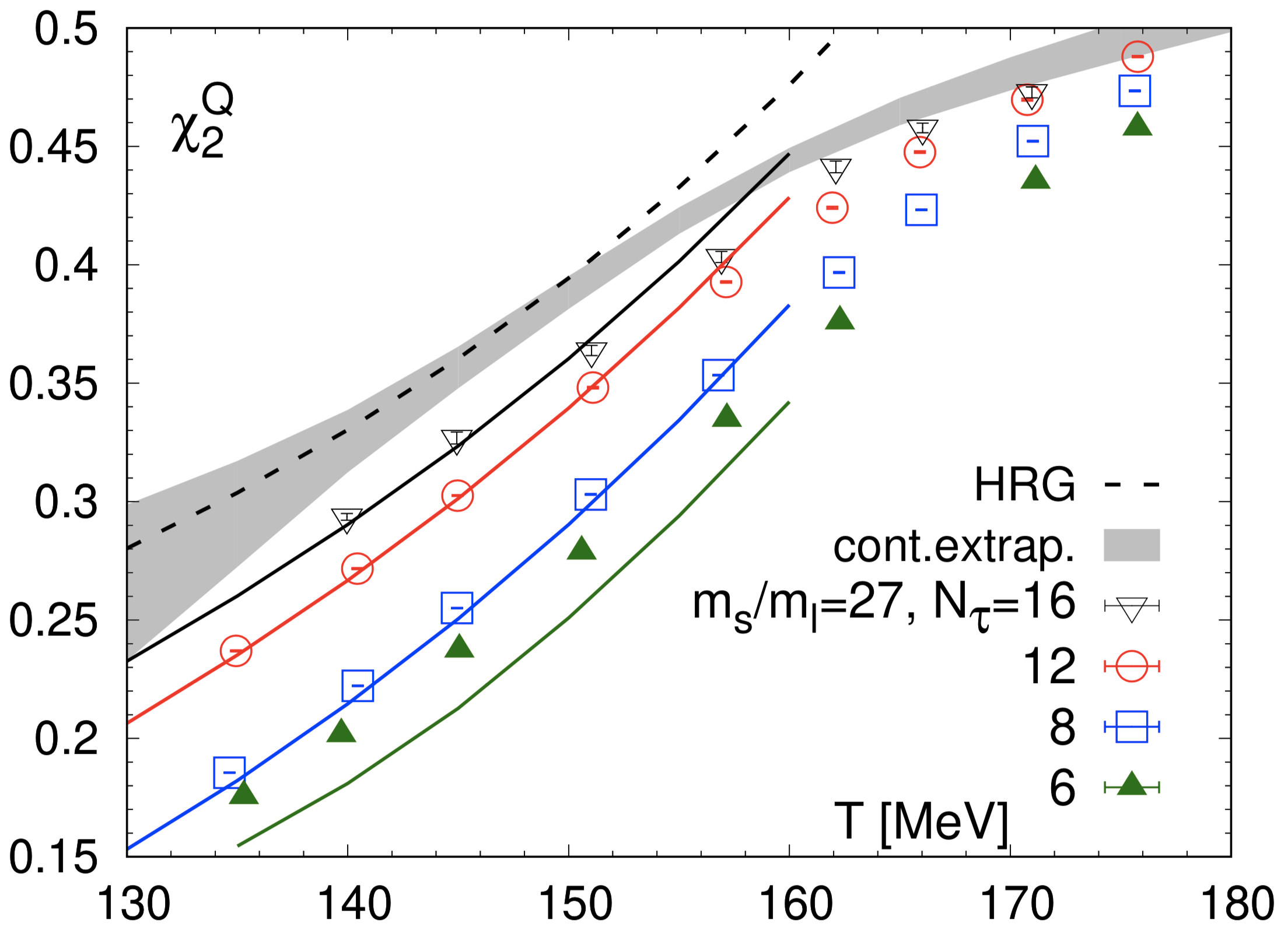}
  \vspace{-1ex}
  \caption{Cut-off dependence of net electric charge fluctuations and continuum
  extrapolation (gray band). Lines show HRG model calculations taking into account
  modifications of the pion masses in calculations with staggered fermions (taste
  symmetry violations). }
  \vspace{-2ex}

\end{wrapfigure}

In the experimental searches for the QCD critical point measurements of higher order
cumulants of net conserved charge fluctuations play important role. The kurtosis and
skewness of net proton-number (as a proxy for net baryon-number), net kaon number (as
a proxy for net strangeness) and net electric charge fluctuations are being measured.
Among these the kurtosis and skewness of net proton-number fluctuations show the
strongest dependence on the beam energy and, hence, on $\mu_B$. For
$\sqrt{s_{_{NN}}}\gtrsim19$~GeV the systematic of kurtosis and skewness can be
reproduced in lattice QCD calculations of net baryon-number fluctuations. It could be
shown that deviations from the simple Skellam distribution, as expected in hadron
resonance gas (HRG)  model calculations, is (i) negative, (ii) increases in magnitude
with with increasing $\mu_B$, and (iii) is about a factor three larger for the kurtosis than the
skewness.

Similar calculations for strangeness and electric charge fluctuations do not yet
exist, but need to be done. In particular, getting quantitative control over the net
electric charge fluctuations is important as these can be compared directly to
experimental results. The calculation of higher order cumulants of electric charge
fluctuations in lattice QCD is challenging for several reasons. They are dominated by
contributions from pions. This introduces a large correlation length $\xi\sim
1/m_\pi$, and the results are very sensitive to finite volume effects. Moreover,
continuum extrapolations are difficult as the pion sector is strongly distorted in
lattice QCD calculations with staggered fermions (because of taste symmetry violations), and
calculations in other discretization schemes are much more demanding in terms of computational
resources. Well-controlled results exist at present for the quadratic fluctuations
($\chi_2^Q$) of net electric charges at $\mu_B=0$. \emph{Calculations on large
spatial lattices and closer to the continuum limit are needed to arrive at controlled
continuum extrapolations for higher moments of net electric charge fluctuations.}

  \begin{wrapfigure}{l}{0.52\textwidth}
    \centering

  \vspace{-2ex}
  \includegraphics[width=0.5\textwidth, height=0.25\textheight]{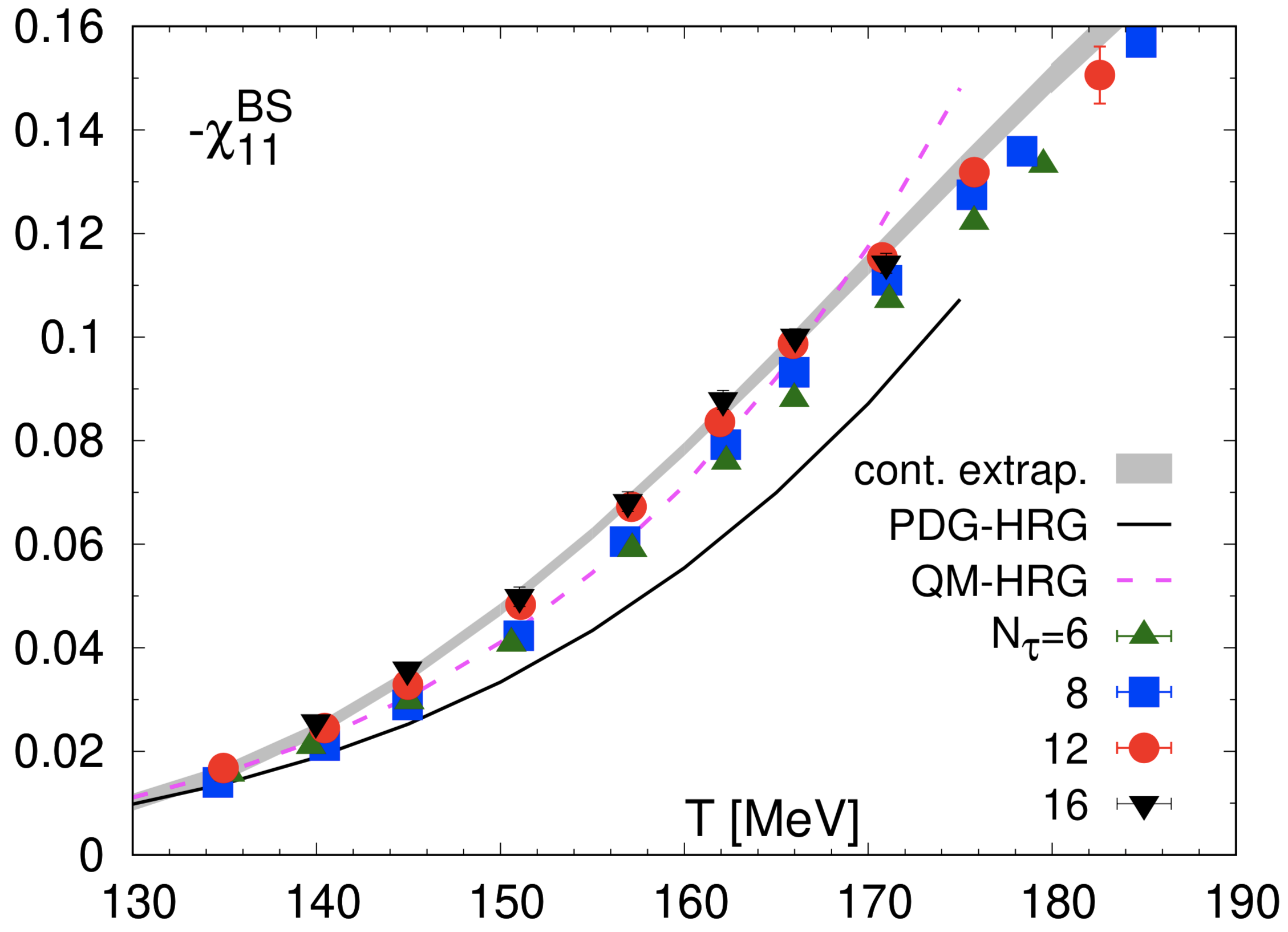}
  \vspace{-1ex}
  \caption{Baryon-number---strangeness correlations at $\mu_B=0$. }
  \vspace{-2ex}

\end{wrapfigure}

Although the HRG model provides a rather good description of many features found in
lattice QCD calculations below the crossover temperature, there are also many
striking deviations from simple HRG model predictions. For instance, in the
strangeness sector quadratic and quartic fluctuations are enhanced over those of
simple HRG model predictions, which dominantly arise from larger
baryon-number---strangeness correlations. This has been attributed to contributions
from additional baryon resonances that are predicted to exist in quark model (QM-HRG)
calculations, but have not yet been observed experimentally (PDG-HRG). They may be
quite broad and their contribution may be taken care of in modified HRG models that
take into account various decay channels of such unstable resonances through a virial
expansion \cite{Fernandez-Ramirez:2018vzu}. In order to better quantify deviations of lattice QCD in various
fluctuation and correlation observables from simple HRG model calculations, and
control additional parameters that enter calculations with extended HRG models
detailed analyses of higher order cumulants are needed. Correlations between
fluctuations in different conserved charge sectors, \textit{e.g.}, the correlation
between baryon-number fluctuations and those of strangeness or electric charge, are
currently being analyzed experimentally at RHIC and LHC. These correlations will also
be studied in the BES-II at RHIC. \emph{In order to calculate charge correlations at
non-zero $\mu_B$, again, accurate results on higher order cumulants are needed.}


\section{Probing QGP with heavy quarks}

Hadrons containing heavy quarks provide an important probe of hot and dense matter
created in heavy ion collisions. For example, quarkonia, mesons composed of a heavy
quark and anti-quark have been proposed as the probe of the  temperature of the
produced medium~\cite{Matsui:1986dk}. The presence of the  hot deconfined medium
weakens the binding effects between the heavy quarks, eventually leading to the
dissolution of the quarkonia. The spectra and angular distributions of the hadrons
with single heavy quark can be used to study the relaxation time scales of quark
gluon plasma~\cite{Moore:2004tg}. More precisely, these observables are sensitive to
the heavy quark  diffusion constant $D \sim M/T t_{relax}$, with $t_{relax}$ being
the typical relaxation time scale of the medium and $M$ being the heavy quark mass.
Together the above probes are commonly referred to as the heavy flavor probes. There
has been a large experimental effort at RHIC and LHC on the studies of heavy flavor
probes. Future sPHENIX experiment at RHIC and ALICE upgrades largely target the
physics of heavy flavor probes.

\begin{wrapfigure}{l}{0.52\textwidth}
  \centering

    \vspace{-2ex}
    \includegraphics[width=0.5\textwidth, height=0.25\textheight]{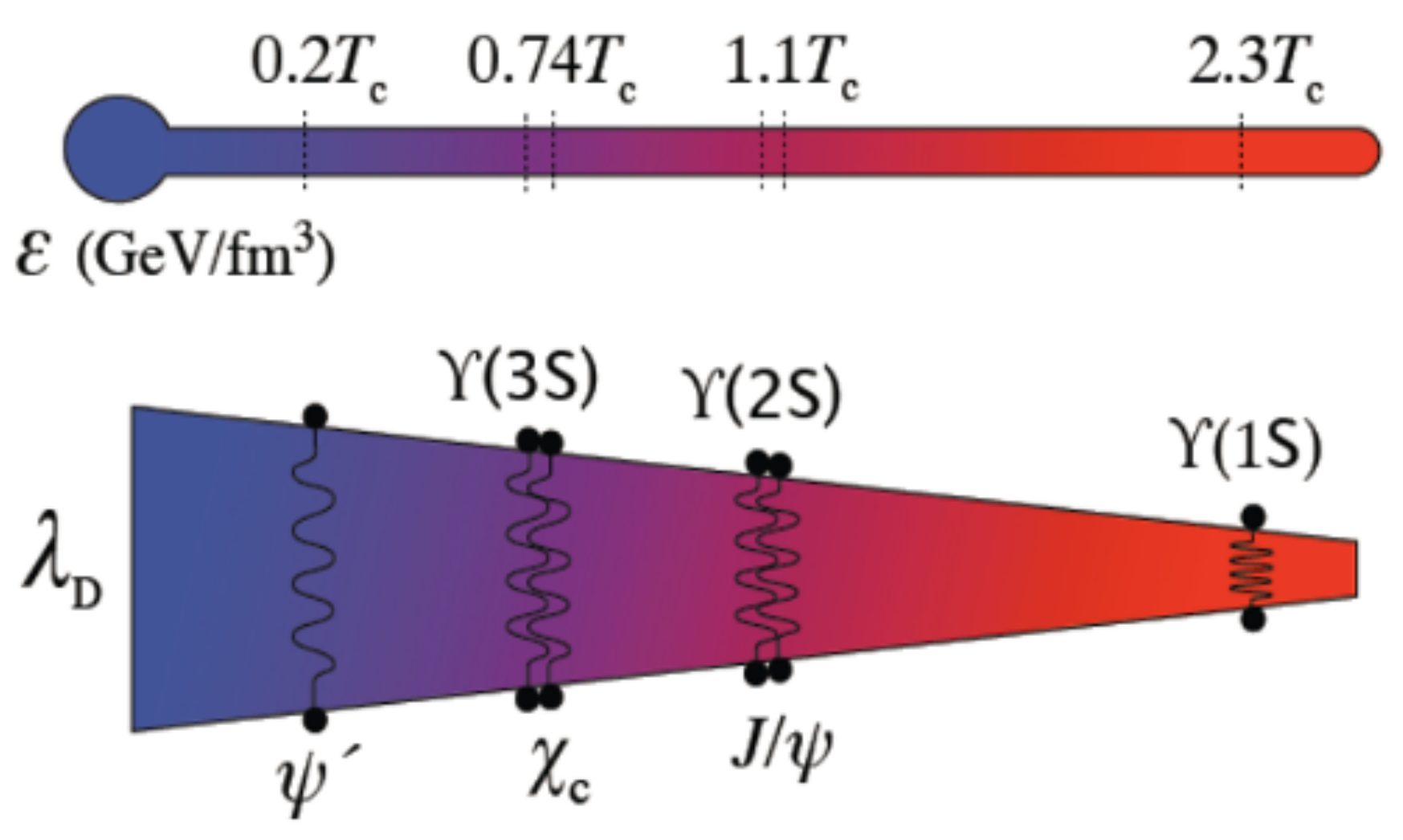}
    \vspace{-1ex}
    \caption{Illustration of how sequential melting of quarkonia of different
    sizes, immersed in QGP, can probe QGP at different length and energy scales.}
    \vspace{-1ex}

\end{wrapfigure}

In-medium properties and/or dissolution of heavy flavor hadrons as well as the heavy
quark diffusion constant are encoded in the spectral functions (see e.g.
Ref. \cite{Mocsy:2013syh} for a recent review). The properties of bound
states  are encoded in
peak-like structures in the spectral functions at values of frequency $\omega$ of the
order of the heavy quark mass. Heavy quark diffusion constant $D$ is encoded in the
behavior of the spectral function for $\omega \simeq 0$. In this region the spectral
function has a peak, often called the transport peak. The width of the transport peak
is proportional to $T/(M D)$ \cite{Petreczky:2005nh} and, thus, is very small.
At frequencies significantly above the bound state peaks the spectral function is featureless,
and this part of the spectral function is referred to as continuum. At sufficiently high
temperatures the bound state peaks  will broaden and disappear and the spectral function for $\omega$ larger
than the quark mass will be described by the continuum, i.e. we will see the melting of
the heavy quark bound states. It is expected that excited quarkonium states will "melt" at smaller temperatures
than the more tightly bound ground state. This is often referred to as the sequential quarkonium melting.

While
there is a direct relation between the spectral function and the Euclidean time
correlation function appropriate reconstruction of the former is very challenging as the
correlation function is available at a discrete set of points and has statistical
errors (see discussion in Ref. \cite{Mocsy:2013syh}). Moreover, the extent of the imaginary time direction is proportional to
$1/T$, and, thus, becomes small at high temperature $T$. One can consider spatial
correlation functions of meson operators, which are not restricted to small
separation, but the relation between the correlation functions and spectral functions
is less direct \cite{Karsch:2012na,Bazavov:2014cta}. Finally, we should point out that discretization effects due to the
heavy quark mass could be also large, especially for the bottom quark.

\begin{wrapfigure}{l}{0.52\textwidth}
  \centering

    \vspace{-4ex}
    \includegraphics[width=0.5\textwidth, height=0.25\textheight]{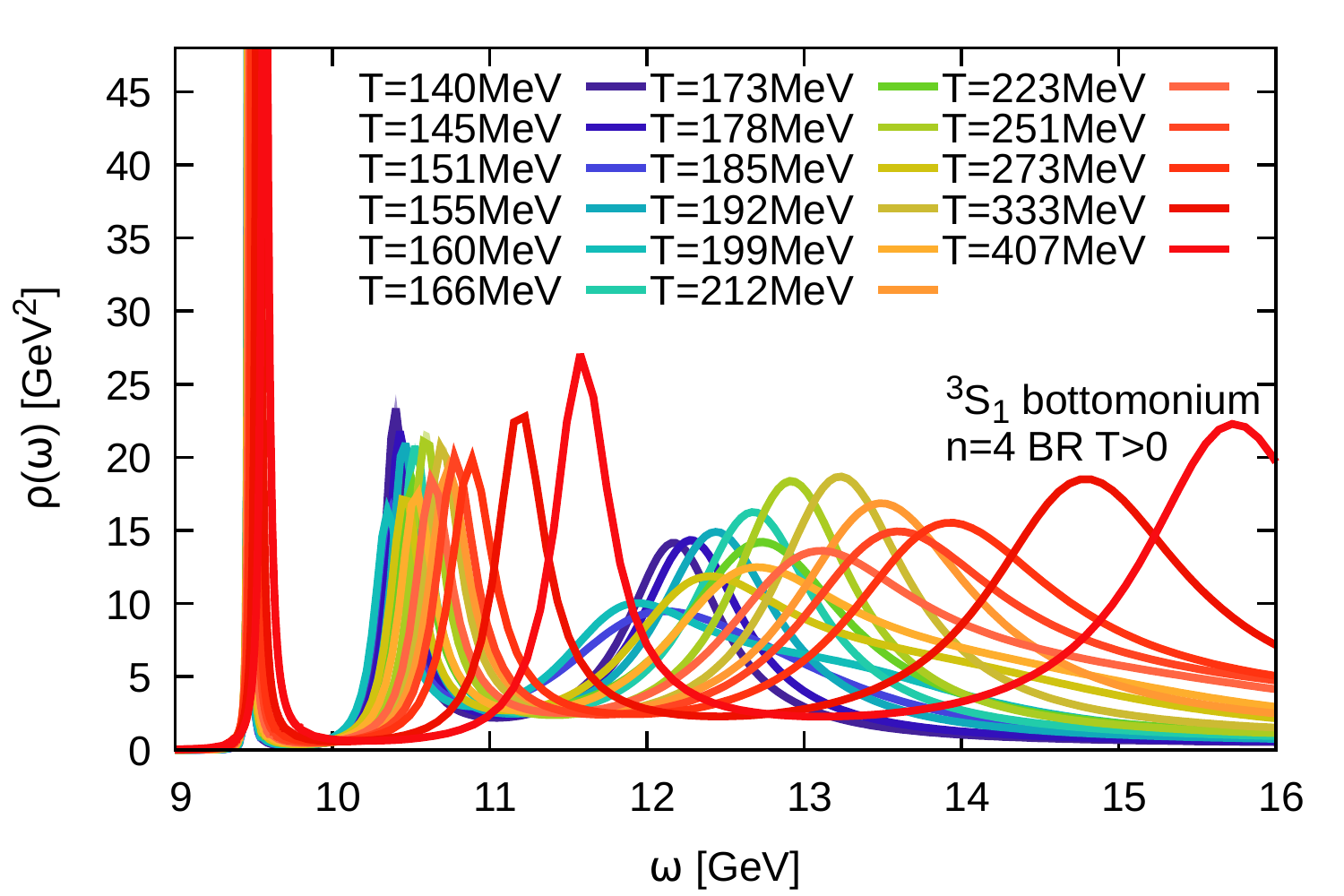}
    \vspace{-1ex}
    \caption{ The $\Upsilon$ spectral functions for different temperatures from
    lattice NRQCD calculations \cite{Kim:2018yhk}.}
    \vspace{-2ex}

\label{fig:bb_spf}
\end{wrapfigure}

The problems discussed above limited our ability to obtain reliable results on
in-medium quarkonium properties. We could use the heavy quark mass to our advantage
and combine lattice QCD with an effective field theory (EFT) approach. Integrating
out the heavy quark mass scale leads to an EFT called non-relativistic QCD (NRQCD),
where the heavy quarks are represented by Pauli spinors and the creation of heavy
quarks is encoded in higher dimensional operators~\cite{Caswell:1985ui}. Because the
scale associated with the heavy quark mass has been integrated out there are no
discretization errors associated with the heavy quark mass. The maximal time extent
in this formulation is $1/T$, which is twice larger than in the standard relativistic
approach to heavy quarks.  The high energy part of the spectral function is also
smaller. As the result, the Euclidean correlation functions in this approach are more
sensitive to the in-medium quarkonium properties. Lattice QCD studies of quarkonium
spectral functions in this approach have been reported using isotropic lattices,
\textit{i.e.}, lattices with the same lattice spacing in the temporal and spatial
directions~\cite{Kim:2014iga, Kim:2018yhk}, and also using anisotropic lattices,
where the lattice spacing in time is smaller than the spatial lattice
spacing~\cite{Aarts:2014cda, Aarts:2013kaa, Aarts:2012ka, Aarts:2011sm,
Aarts:2011sm}. The latter calculations did not reach the physical quark masses, while
the former are limited to smaller lattices with  temporal extents $N_{\tau}=12$. Both
calculations, however, show that $\Upsilon(1S)$ state can survive in the deconfined
matter up to temperatures as high as $400$ MeV, with only small medium modifications. The
spectral functions reconstructed from $N_{\tau}=12$ lattice
calculations~\cite{Kim:2018yhk} are shown in Fig.~\ref{fig:bb_spf}. One can clearly
see the first peak corresponding to $\Upsilon(1S)$, with relatively little
temperature dependence. Most of the studies of quarkonium spectral functions, including
the ones mentioned above rely on correlation functions of point meson operators, i.e.
local quark bilinears. These correlators are dominated by the continuum part of the spectral
function at high temperatures and thus have limited sensitivity to the in-medium quarkonium
properties. This is especially true for the excited bottomonium states \cite{Kim:2018yhk}.
Using correlators of extended meson operators it is possible to get more sensitivity
to the in-medium quarkonium properties as the relative contribution of the high frequency part of
the spectral function will be smaller in this case.
\emph{In future it will be important to extend the
calculations to physical quark masses using anisotropic lattices, so that quarkonium
correlators with large number of data points in the time direction are available.
Furthermore, correlation functions of extended meson operators should be studied in order
to extract information about the in-medium properties of excited quarkonium states and confirm
the expected pattern of sequential quarkonium melting.
Also, one should consider lattice calculations on isotropic lattices much closer to
the continuum limit~\cite{Burnier:2017bod}.}

If we integrate out the energy scale associated with the inverse size of the
quarkonium we get another EFT, the potential non-relativistic QCD (pNRQCD). The
degrees of freedom in this EFT are the singlet and octet static meson fields, and the
static quark anti-quark potential enters as the parameter in the Lagrangian of this
EFT~\cite{Brambilla:1999xf}. In this case, the  static quark anti-quark potential
entering the EFT Lagrangian not only becomes temperature dependent, but also turns
out to be  complex~\cite{Brambilla:2008cx}.  The complex potential can be calculated
in weak coupling approach if the temperature is sufficiently high. When the binding
energy is the smallest scale in the problem all higher energy scales can be
integrated out and all the medium effects can be encoded in the temperature dependent
potential \cite{Petreczky:2010tk}. Furthermore, the potential can be calculated on
the lattice by considering temporal Wilson loops. Thus, the problem of the in-medium
properties is reduced to the calculation of Wilson loops at non-zero temperature and
extracting the complex potential from them.  The study of quarkonium properties in
pNRQCD is very important as it provides a link between QCD and dynamical models of
quarkonium production in heavy ion collisions, see \textit{e.g.}
Refs.~\cite{Brambilla:2017zei,Brambilla:2016wgg}. Lattice calculations of the complex
potential have been carried out both in quenched~\cite{Rothkopf:2011db} and 2+1
flavor QCD~\cite{Burnier:2014ssa,Bazavov:2014kva}. In the latter case the calculations were
limited to temporal extent $N_{\tau}=12$, and
un-physically heavy dynamical light quark masses. \emph{It will be important in the
future to extend these calculations to physical quark masses and larger $N_{\tau}$.}

As mentioned above the heavy quark diffusion coefficient is related to the behavior
of the spectral functions at zero energy. The spectral function in this energy region
has a peak, called the transport peak, which is a Lorentzian in the heavy quark
limit, $\sigma_{trans}(\omega)=\chi_q \eta/(\omega^2+\eta^2)$, $\chi_q$ being the
heavy quark number susceptibility~\cite{Petreczky:2005nh}. For heavy quarks
$\eta=T/(M D) \ll 1$, \textit{i.e.}, the transport peak is very narrow. Therefore, an
accurate determination of its width and thus the diffusion constant $D$ is extremely
challenging~\cite{Petreczky:2005nh}. However, we could take advantage of this feature
and use the Heavy Quark Effective Theory (HEQT), where the heavy quark degrees of
freedom are integrated out~\cite{CaronHuot:2009uh}. In this approach, one calculates
the correlation function of chromo-electric field strength, which gives the momentum
space diffusion coefficient $\kappa$ as the $\omega \rightarrow 0$ limit of the
corresponding spectral function. The spectral function in this case does not have a
peak at $\omega \sim 0$ and is smoothly connected to the large $\omega$ region.
Instead of determining the width of the transport peak one needs to determine the
intercept at $\omega=0$, which is much easier. Calculations along these lines have
been performed in quenched QCD~\cite{Banerjee:2011ra, Francis:2015daa} and resulted
in the value $\kappa = (1.8 - 3.4)T^3$. The reason for performing this calculation in
quenched QCD is that the correlator of chromo-electric field strength is very noisy,
and a novel multilevel algorithm is needed to obtain a good signal to noise
ratio~\cite{Luscher:2001up,Meyer:2002cd}. This algorithm is presently only available for quenched
QCD. \emph{There are two possible ways  to extend the existing calculations. One
should consider higher temperatures, where weak coupling calculations are expected to
work and comparisons between the lattice and weak coupling calculations are possible.
This will be very important for validating the procedure of extracting the spectral
functions from the lattice data. Second, one needs to develop algorithms in full QCD,
which can deal with the noise problem.
This remains a very challenging task despite recent progress \cite{Ce:2016idq}.
}

So far, we have discussed quarkonium in-medium properties. In-medium properties of
open heavy flavor hadrons are far less explored on the lattice. Attempts to study D
mesons at non-zero temperature have been presented in Ref.~\cite{Kelly:2018hsi}. An
alternative way to study in-medium hadron properties is to consider spatial
correlation functions~\cite{Karsch:2012na, Bazavov:2014cta}. The spatial correlation
functions are not limited to $1/T$ and, therefore, are more sensitive to the
in-medium properties of hadrons. However, the relation of the spatial hadron
correlation functions to the spectral functions is more complicated, involving a
double integral transformation~\cite{Karsch:2012na}. At low and very high
temperatures the relation is simple and can be used to constrain the in-medium
properties of heavy flavor hadrons.

Yet another way to obtain insights into the in-medium properties of open charm
hadrons is to study charm fluctuations and correlations~\cite{Bazavov:2014yba}. The
corresponding lattice calculations indicate that charm hadronic excitations may exist
above $T_c$~\cite{Mukherjee:2015mxc}. Fluctuations of charm and charm-baryon
correlations are also interesting from the point of view of providing information
about the spectrum of charm hadrons in the vacuum. The lattice calculations
indicate that there are additional charm baryons not listed by the Particle Data Group
(PDG) but expected based on quark model calculations~\cite{Bazavov:2014yba}.
Current calculations have been performed on coarse lattices and/or unphysical quark
masses. \emph{It will be important to extend these calculation to finer lattices and
physical quark masses.}

\section{Nature of QCD phase transition}

Although, by now, it is well-established that the transition from the low temperature
hadronic phase of QCD to the asymptotically free quark-gluon phase is not a genuine
phase transition, but a smooth crossover, it is expected that this crossover is
sensitive to properties of strong-interaction physics that are described by a true
phase transition in the limit of vanishing up and down quark masses, \textit{i.e.},
in the chiral limit. Understanding the properties of strong-interaction physics at
non-zero temperature and baryon chemical potential continues to be an extremely
active field of research both experimentally and theoretically. It is, thus, of
utmost importance to lay the ground for this research by firmly establishing the
phase structure of QCD in the chiral limit. By doing so, we will be able to quantify
to what extent the non-analytic features QCD that dominate the physics of strongly
interacting matter in the chiral limit contribute to the non-perturbative properties
of strongly interacting matter observed in nature.

\begin{wrapfigure}{l}{0.52\textwidth}
  \centering

 \vspace{-2ex}
  \includegraphics[width=0.5\textwidth, height=0.25\textheight]{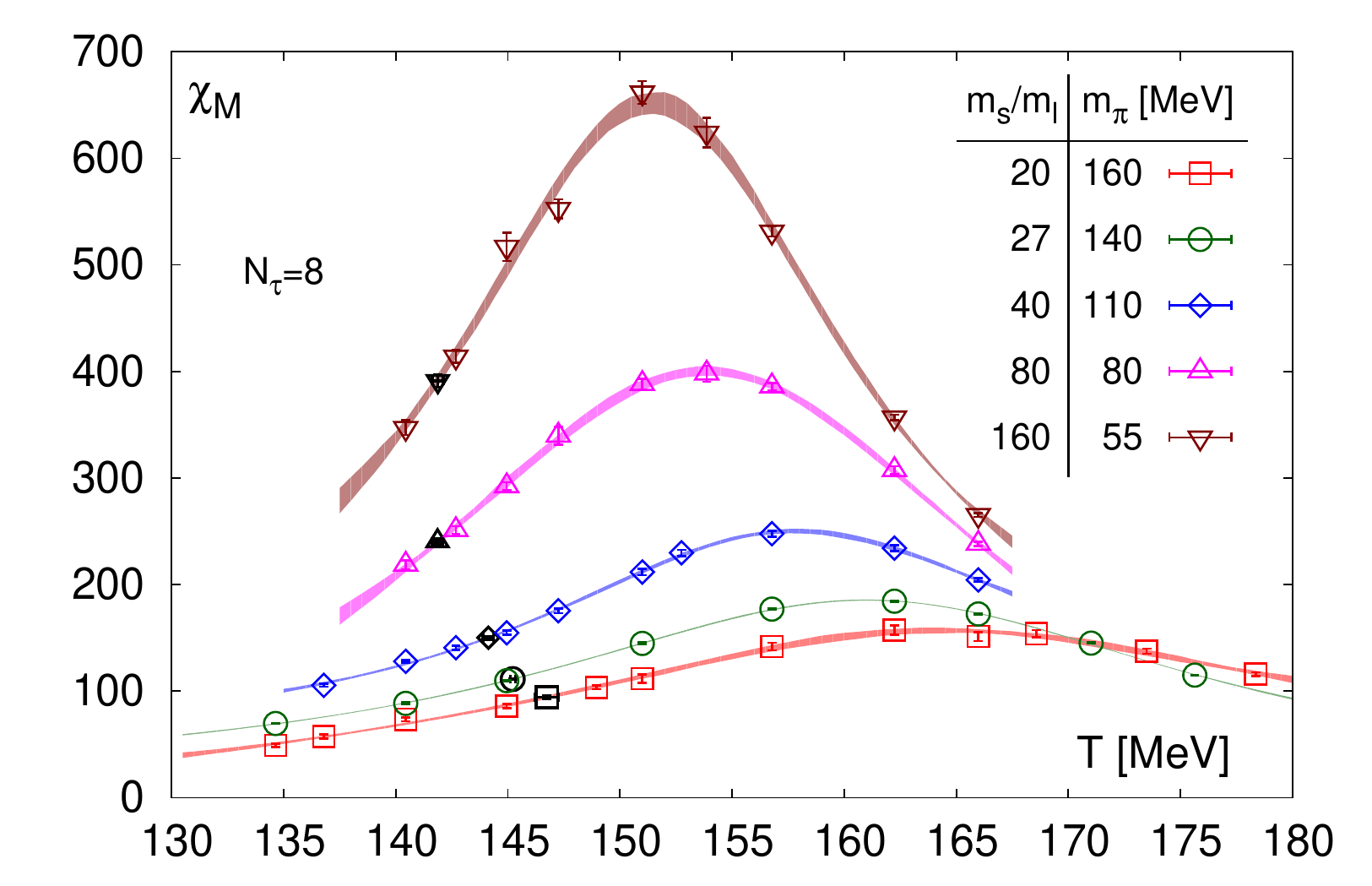}
  \vspace{-1ex}
  \caption{The chiral susceptibility of (2+1)-flavor QCD on lattices with temporal
  extents $N_\tau=8$ and for various values of the light quark masses~\cite{Ding:2019prx}.}
  \vspace{-1ex}

\end{wrapfigure}

A first obvious question that needs to be clarified is how the QCD transition
temperature depends on the value of the light quark masses. What is the critical
temperature in the chiral limit, and is this transition a $2^{nd}$ order transition
or does it turn into a $1^{st}$ order transition at some critical value of the light
quark masses ? These questions are closely related to the question of
(non-) existence of the QCD critical end-point in $(T-\mu_B)$-plane. Unlike earlier studies of the nature of the transition in the chiral
limit, that used unimproved staggered or Wilson fermion discretization schemes,
current studies of the order of the chiral transition, performed with improved
actions~\cite{Bazavov:2017xul,Ding:2018auz,Ding:2019prx}, do not find any hint for a first order
transition down to Goldstone pion masses as small as $55$~MeV. \emph{These studies,
currently, are performed on moderately sized lattices. Calculations closer to the
chiral limit and proper continuum extrapolations are needed to confirm these
results.}

First attempts to determine the chiral phase transition temperature in the continuum
limit have been undertaken recently. Our current understanding is that the chiral
phase transition in QCD with two mass-less quarks and a physical strange quark occurs
at $T_c = 132^{+3}_{-6}$~MeV~\cite{Ding:2019prx}, \textit{i.e.}, at a temperature
about 25~MeV smaller than the QCD transition
temperature~\cite{Steinbrecher:2018phh}.

The nature of the chiral phase transition also crucially depends on the fate of the
axial $U_A(1)$ symmetry at high temperatures. This was understood early on by
Pisarski and Wilczek~\cite{Pisarski:1983ms}. They speculated that the QCD phase
transition might be first order, if the chiral anomaly would get 'effectively
restored' at the phase transition temperature for chiral flavor symmetry restoration.
However, recently it has been shown that a universality class for $O(2)\times O(N)$
models in three dimensions exists~\cite{Calabrese:2004nt,Pelissetto:2013hqa}. This
makes it possible that the QCD phase transition is second order, and at the same
time, may lead to an effective restoration of the axial $U_A(1)$ symmetry.

A crucial indicator for the chiral $U_A(1)$ symmetry restoration is the structure of
the low-lying eigenvalue spectrum of the Dirac operator. In order for $U_A(1)$ to
remain broken a non-zero density of near-zero modes needs to be present on finite
lattices, which eventually will give rise to a non-zero density of zero modes that
can be responsible for $U_A(1)$ symmetry breaking. In recent calculations within
chiral fermion discretization schemes and physical values of the light quark masses
arguments have been put forward in favor of $U_A(1)$ symmetry restoration close to
the flavor symmetry restoring transition~\cite{Suzuki:2017ifu}. However, these
calculations still are done on rather small lattices, which makes it difficult to
eliminate the influence of finite-volume effects on the relevant small Dirac
eigenvalues. Moreover, due to the large difference between the crossover temperature
at physical values of the quark masses and the chiral phase transition temperature it
is obvious that these calculations, at present, only allow to conclude that $U_A(1)$
symmetry breaking effects become small at temperatures $T\ge 1.2T_c$. \emph{More
detailed studies close to the chiral limit and on large lattices are needed to arrive
at definitive conclusions on the role of $U_A(1)$ symmetry breaking close to the
chiral phase transition.}


\section{Electromagnetic probes of QGP}

QGP is a thermal medium consisting of electrically charged quarks and, hence,
naturally emits photons. Due to the smallness of the electromagnetic coupling and
limited extent of the medium created during collisions of heavy ions, photons emitted
inside QGP escape the medium without subsequent interactions. Also, QGP-emitted
virtual photons decay into lepton pairs (di-leptons) and escape the medium without
further interaction.  Thus, photons and di-leptons provide valuable information
regarding properties of hot-dense QCD matter~\cite{Rapp:2016xzw}, and are
experimentally sought after observables at RHIC and  LHC~\cite{Campbell:2017kbo}.
Furthermore, the strengths of the experimental signals of the chiral magnetic effect
sensitively depend on how long the magnetic field, produced at very early times,
lasts inside expanding QGP~\cite{Kharzeev:2015znc}; the lifetime of the magnetic
field trapped inside QGP is entirely governed by the value of the electrical
conductivity of QGP~\cite{McLerran:2013hla}.

Lattice QCD calculations of di-lepton rates, photon emissivity and electrical
conductivity rely upon reliable extraction of the spectral function from Euclidean
correlation function of the vector current. These studies demand very precise
calculations of the vector current two-point correlation function on lattices having
very large temporal extents. Present lattice QCD results are either based on quenched
approximation~\cite{Ding:2016hua, Ghiglieri:2016tvj}, or carried out on small
lattices with un-physically heavy dynamical quarks~\cite{Aarts:2014nba,
Brandt:2017vgl}. \emph{To have an  impact on the experiments, in future, lattice QCD
calculations of electromagnetic probes of QGP need to be carried out on large
lattices with physical dynamical quarks.}


\section{Exploring jet energy loss and viscosities}

Probing properties of QGP through detailed studies of jet quenching, \textit{i.e.},
the energy loss of a fast moving parton inside QGP,  is one of the key components of
the future sPHENIX experiment at RHIC, as well as the LHC heavy-ion program. Jet
quenching in QGP is characterized in terms of a quantity called $\hat q$ that
measures the momentum transfer squared per unit time. A fully non-perturbative
estimate of $\hat q$ is highly desired. As the mechanism of jet quenching involves
dynamics of a fast moving parton on the light cone, a direct lattice QCD-based
determination of $\hat q$ is an extremely challenging problem. However, at
temperatures significantly higher than the QCD crossover temperature  one can use an
EFT-based approach to address this problem~\cite{CaronHuot:2008ni, Benzke:2012sz}. In
this case,  one can use the 3-dimensional EFT, the electrostatic QCD (EQCD), to
calculate $\hat q$ by separating the perturbative and non-perturbative parts of the
calculation. One can solve EQCD on the lattice and determine the non-perturbative
part of $\hat q$. Calculations along these lines have been presented in
Ref.~\cite{Panero:2013pla}.  A different approach to calculate the jet quenching
parameter $\hat q$ based on the field strength correlator on the light cone, was
suggested in Refs.~\cite{Kumar:2018cgf, Majumder:2012sh}. In lattice setup this
reduces to the calculations of the expectation value of a local operator. It is not
clear, however, to what extent local operators in Euclidean space time can
approximate the physics on the light cone.

The calculations of the shear and bulk viscosities in lattice QCD are also extremely
challenging. There are at least two reasons for this. First, the viscosities are
defined in terms of correlators of energy-momentum stress tensor, which involves
gluonic quantities. Gluonic quantities are very noisy when evaluated on the lattice.
Second, the correlators of energy-momentum stress tensor are dominated by the high
frequency modes, and the corresponding spectral functions are proportional to
$\omega^4$ at large frequencies. As the result, the corresponding correlators have
little sensitivity to the transport peak. Viscosities that are different by an order
of magnitude will lead to changes in the correlation functions by less than a
$1\%$~\cite{Pasztor:2018yae}. To deal with the first problem multi-level algorithm
should be used~\cite{Luscher:2001up,Meyer:2002cd}. This, presently, limits the calculations to pure
$SU(3)$ gauge theory. To deal with the second problem on can use Ward identities to
obtain correlators that correspond to spectral functions behaving like $\omega^2$ at
large frequencies~\cite{Meyer:2008gt, Meyer:2011gj}.  Using these tricks and
anisotropic lattices the best determination for shear viscosity to entropy ratio
gives $\eta/s=0.17(2)$ for pure $SU(3$) gauge theory at
$T=1.5T_c$~\cite{Pasztor:2018yae}. This seems to confirm the strongly coupled nature
of QGP, though the systematic uncertainties due to modeling the spectral functions
may not be fully understood. Another study finds $\eta/s=0.27(7)$ for
$1.5T_c$~\cite{Astrakhantsev:2017nrs}, which is higher than the above estimate. Thus,
a better understanding of the analytic form of the spectral function is needed.
Analytic calculations on the weak coupling side may help in this
respect~\cite{Hong:2010at}.

Given the extremely challenging nature of both these problems, presently, it is
unclear whether fully-controlled lattice calculations, with light dynamical fermions,
of $\hat q$ and QGP viscosities can be achieved in the near future. However, albeit
their challenges, these problems present immense opportunities for hot-dense lattice
QCD to impact the explorations of QGP properties. Thus, it will be very important to
explore all possible new avenues, both algorithmic and formalism wise, to make
significant progress in addressing these issues on the lattice.


\section{Computational challenges}

In the last decade, significant progresses toward quantifying the features of the hot
and dense strong-interaction matter have been made in the \textit{ab-initio} lattice
QCD calculations, as has been outlined in the previous sections. It became possible
with the continued DOE support for computing hardware, as well as software
development efforts. Currently members of USQCD utilize the dedicated USQCD hardware
funded by DOE and the large-scale computational resources available through the
INCITE and ALCC programs.

However, many challenges still remain, which require significant algorithmic
developments and computational power to address them. The need for software
development for future architectures has been recognized by DOE through the
continuation of funding of the Nuclear Physics SciDAC-4, as well as the Exascale
Computing Program. Based on the existing software, available expertise, ongoing
software development efforts and the projected computational resources that are going
to be available, we outline below the computational challenges in relation to the
hot-dense lattice QCD physics program.

At present, there is no direct method to carry out Monte Carlo simulations of QCD at
$\mu_B>0$ due to the sign problem. The way to proceed is to expand the pressure in
$\mu_B/T$ and calculate the physical observables as Taylor expansions in this
quantity. In practice, this requires calculating operators of high order, which are
noisy and require very large statistics. The traces of operators are estimated
stochastically and, thus, one repeatedly solves the Dirac equation on the same
gauge field configuration with many, order thousands, different sources. This problem
requires high-capacity computing, since the work can be split among many GPUs or
multi-core nodes, requiring communication only among few of them. To control the
approach to the continuum limit one also needs to perform calculations at finer
lattices, \textit{i.e.}, with larger temporal and spatial dimensions.

While for calculations at $\mu_B>0$ the main computational cost is in the number of
measurements per given gauge field configuration, studying the properties of QCD
close to the chiral limit poses a different challenge. The cost of inverting the
Dirac operator is inversely proportional to the square of the lightest quark mass.
Given that the current simulations place the bound on the first-order region of the
Columbia plot\footnote{A two-dimensional diagram that represents the expected order of the finite-temperature transition in QCD as function of the light and strange quark masses, as first appeared in Ref.~\cite{Brown:1990ev}.} at the pion mass around 55~MeV or smaller, one needs to perform
simulations at light quark masses, significantly lower than their physical values. On
top of that, to maintain full control over the finite-volume effects, one needs
larger lattice volumes than in simulations with the physical light quark masses. For
the large lattices the problem is in the high-capability domain, where efficient
inter-node parallelism is needed. The ongoing software development efforts are
towards multi-GPU codes as well as a hybrid MPI-OpenMP model that can make efficient
use of multi-core architectures. Besides that, algorithmic improvements such as
multigrid~\cite{Brannick:2007ue,Babich:2010qb,Brower:2018ymy} are needed. At present,
efficient multigrid algorithm has been designed for Wilson fermions, while it remains
an open question if the same level of efficiency can be achieved for staggered
fermions~\cite{Brower:2018ymy}.

Simulations at fine lattices and low quark masses suffer from critical slowing down,
related to the freezing of the gauge field topology. This area may need further
algorithmic improvements, since subtle properties such as the restoration of the
anomalous $U_A(1)$ symmetry depend on the proper sampling of the topology of the
gauge fields. The symmetry properties of lattice fermions may also be important in
this problem. While Domain Wall Fermions (DWF) are not used in large-scale hot-dense
QCD calculations due to their high computational cost, they possess almost exact
chiral symmetry on the lattice. To fully understand the phenomena associated with the
$U_A(1)$ symmetry above the chiral crossover temperature a program of large-scale DWF
calculations might be needed in future.

The fate of the heavy-quark bound states in Quark-Gluon Plasma has been of
considerable interest since the Matsui-Satz conjecture~\cite{Matsui:1986dk}, that
suppression of $J/\psi$ yields due to the screening effects in the thermal medium may
provide an unambiguous signal of QGP formation. In reality, there are several
competing effects and the overall picture turned out to be more complicated. To fully
address the properties of heavy quarkonia one needs theoretical understanding of the
medium modification of their spectral functions that encode full information about
the states. Since lattice QCD is formulated in the Euclidean space-time formalism,
the real-time properties, such as the spectral functions, are hard to access. In
practice, one computes Euclidean correlation functions, from which the spectral
functions can be reconstructed by solving and inverse problem. This inverse problem
is very ill-posed since the available input information is limited by the number of
lattice points in the temporal direction.

The isotropic (\textit{i.e.} the same lattice spacing in the temporal and spatial
directions) gauge field configurations that will be generated to address the QCD at
$\mu_B>0$ and close to the chiral limit are limited to the temporal extent of
$N_\tau=24$ and maybe $N_\tau=32$ in the longer term. For robust spectral function
reconstruction one needs lattices with $N_\tau$ of order a hundred. Thus, a dedicated
program of computations on anisotropic lattices is needed. With anisotropy factor of
$6$ one can reach finite-temperature lattices $32^3\times48$ to $64^3\times96$ that
correspond to temperatures of $2-3T_c$.

Apart from generating dedicated finite-temperature lattices with large $N_\tau$,
designing new reconstruction algorithms is also of importance. The Bayesian
techniques such as the Maximum Entropy Method (MEM)~\cite{Asakawa:2000tr} are often
used. Recently a modification of MEM has been introduced in~\cite{Burnier:2013nla},
which however still has some deficiencies, such as ringing, that may produce false
peaks in the spectral functions. Further improvements, perhaps based on the recent
progress in the field of machine learning, may be helpful in this area.

From the methodological point of view, the problem of heavy quarks in QGP is deeply
related to first-principle calculations of transport properties of QGP. Calculation
of shear and bulk viscosities, photon emissivity, electrical conductivity, and all
such tranport properties depend on fully-controlled reliable extraction of spectral
functions from the corresponding Euclidean correlation functions. The viscosities
represent the grand challenge since they come from the noisiest  and the least
understood channel, related to the correlators of the energy-momentum stress tensor. Thus,
developing the inverse problem methodology for heavy quarks and electromagnetic
probes will pave the way for first-principle shear and bulk viscosity calculations.
In the long run, the latter will require development of new methods beyond what is
currently available on the lattice. Possible collaboration with data scientists, that
deal with various facets of the inverse problem spanning across many scientific
domains, may prove useful for the lattice community. At the same time, continued
focused efforts on algorithmic improvements to produce anisotropic lattices with
large temporal extents will ensure near-term availability of high-quality
next-generation data sets of gauge field samples.

Steady progress of lattice QCD calculations in general and at finite temperature and
density in particular in the last several decades has been achieved by constant
improvement of the theoretical understanding of the underlying quantum field theory
as well as development of new efficient algorithms for simulating the theory on the
lattice. Often times, this happens by trial and error and it is thus hard to predict
which particular direction may lead to a breakthrough. It is thus important apart
from maintaining the well-established and planned-ahead research program to leave
some room for trying new ideas. Our discussion will be incomplete without mentioning
some high-risk high-return directions, which, if progress is achieved, will
significantly impact the class of problems that can be addressed in hot-dense lattice
QCD.

The long-standing sign problem is still drawing attention in the lattice community.
One possible angle to attack the problem is to develop a formalism that allows for
simulating arbitrary, including complex, actions. This is being pursued with the
complex Langevin and the Lefshetz thimble methods. Selected recent work can be found
in \cite{Aarts:2016mso,Aarts:2016bdr,Alexandru:2018ngw,Bluecher:2018sgj}. Another
angle is to rely on the Taylor expansion and/or the imaginary chemical potential
methods, but to find ways to speed them up dramatically, so that much higher orders
in the expansion become achievable~\cite{deForcrand:2017cja,deForcrand:2018zll}. In
the inverse problem class that spans from in-medium properties of heavy-flavor states
to transport properties of QGP an interesting possibility is an attempt to perform
simulations in the real-time Schwinger-Keldysh formalism
\cite{Pawlowski:2016eck,Alexandru:2016gsd,Alexandru:2017lqr}.

\bibliographystyle{apsrev4-1}
\bibliography{ref}

\end{document}